\documentstyle[12pt]{article}

\begin{document}

\setcounter{page}{1}

\pagestyle{plain} \vspace{1cm}
\begin{center}
\Large{\bf Embedding of FRW Cosmology in DGP Scenario with a
Non-Minimally Coupled Scalar Field on the Brane}\\
\small \vspace{1cm}
{\bf Kourosh Nozari$^{a,b}$}\\
\vspace{0.5cm} $^{a}$ {\it Centre for Particle Theory, Durham
University, South Road, Durham DH1 3LE, UK}\\
$^{b}${\it Department of Physics, Faculty of Basic Sciences,
University of Mazandaran,\\
P. O. Box 47416-1467,
Babolsar, IRAN\\
e-mail: kourosh.nozari@durham.ac.uk}

\end{center}
\vspace{1.5cm}
\begin{abstract}
We construct a DGP inspired braneworld scenario where a scalar field
non-minimally coupled to the induced Ricci curvature is present on
the brane. We show that this model allows for an embedding of the
standard Friedmann cosmology in the sense that the cosmological
evolution of the background metric on the brane can be described by
the standard Friedmann equation. The relation between our framework
and the dark-energy formalism is explored.
\end{abstract}
\newpage

\section{Introduction}
Based on light-curves analysis of several hundreds type Ia
supernovae[1,2], observations of the cosmic microwave background
radiation by the WMAP satellite [3] and other CMB-based
experiments[4,5], it has been revealed that our universe is
currently in a period of accelerated expansion. Some authors have
attributed this late-time expansion of the universe to an energy
component referred to as {\it dark energy}. The simplest example in
this regard is the cosmological constant itself which provides a
model of dark energy. However, it is unfavorable since it requires a
huge amount of fine-tuning[6]. Phantom fields[7], quintessence[8]
and modification of gravitational theory itself[9,10] are other
attempts to explain this late time expansion of the universe. In the
spirit of modified gravitational theory, Carroll {\it et al} have
proposed  $R^{-1}$ modification of the usual Einstein-Hilbert
action[11]. It was then shown that this term could give rise to
accelerating solutions of the field equations without dark energy.

On the other hand, theories of extra spatial dimensions, in which
the observed universe is realized as a brane embedded in a higher
dimensional spacetime, have attracted a lot of attention in the last
few years. In this framework, ordinary matters are trapped on the
brane but gravitation propagates through the entire spacetime
[9,12,13]. The cosmological evolution on the brane is given by an
effective Friedmann equation that incorporates the effects of the
bulk in a non-trivial manner[14]. From a cosmological view point,
one of the importance of brane models lies in the fact that they can
provide an alternative scenario to explain the late-time accelerated
expansion of the universe.

Theories with extra dimensions usually yield correct Newtonian limit
at large distances since the gravitational field is quenched on
sub-millimeter transverse scales. This quenching appears either due
to finite extension of the transverse dimensions [12,15] or due to
sub-millimeter transverse curvature scales induced by negative
cosmological constant [13,16-19]. A common feature of these type of
models is that they predict deviations from the usual 4-dimensional
gravity at short distances. The model proposed by Dvali, Gabadadze
and Porrati (DGP) [9] is different in this respect since it predicts
deviations from the standard 4-dimensional gravity even over large
distances. In this scenario, the transition between four and
higher-dimensional gravitational potentials arises due to the
presence of both the brane and bulk Einstein terms in the
action[20]. In this scenario, the existence of a higher dimensional
embedding space allows for the existence of bulk or brane matter
which can certainly influence the cosmological evolution on the
brane. Even if there is no 4-dimensional Einstein-Hilbert term in
the classical theory, such a term should be induced by
loop-corrections from matter fields [21]. Generally one can consider
the effect of an induced gravity term as a quantum correction in any
brane-world scenario.\\
A particular form of bulk or brane matter is a scalar field. Scalar
fields play an important role both in models of the early universe
and late-time acceleration. These scalar fields provide a simple
dynamical model for matter fields in a brane-world model. In the
context of induced gravity corrections, it is then natural to
consider a non-minimal coupling of the scalar field to the intrinsic
(Ricci) curvature on the brane that is a function of the field. The
resulting theory can be thought of as a generalization of
Brans-Dicke type scalar-tensor gravity in a brane-world context.
There are several studies in this direction[22-31]. Some of these
studies are concentrated on the bulk scalar field minimally[22-24]
or non-minimally[25-27] coupled to the bulk Ricci scalar. Some other
authors have studied the minimally[28,29] or non-minimally[30,31]
coupled scalar field to the induced Ricci scalar on the brane.
However, none of these studies have investigated the possibility of
embedding of FRW cosmology in DGP scenario with a 4D non-minimally
coupled scalar field on the brane. The purpose of this letter is to
do this end.

In this letter, in the spirit of DGP inspired gravity, we study the
effect of an induced gravity term which is an arbitrary function of
a scalar field on the brane. We present four-dimensional equations
on a DGP brane with a scalar field non-minimally coupled to the
induced Ricci curvature, embedded in a five-dimensional Minkowski
bulk. This is an extension to a braneworld context of scalar-tensor
(Brans-Dicke) gravity. We show that our model allows for an
embedding of the standard Friedmann cosmology in the sense that the
cosmological evolution of the background metric on the brane can
entirely be described by the standard Friedmann equation plus total
energy conservation on the brane. As original DGP scenario and its
minimal extension which support late-time acceleration of the
universe[20,29], it is easy to show that our model contains this
late time acceleration for a suitable range of non-minimal coupling.
However, with non-minimally coupled scalar field on the brane,
generally it is harder to achieve accelerated expansion[33]. Only
with suitable choice of non-minimal coupling and scalar field
potential one can achieve accelerated expansion in this non-minimal
model.

We use a prime for differentiation with respect to fifth coordinate
except for two cases: $\alpha'\equiv\frac{d\alpha}{d\phi}$ and
$V'\equiv\frac{d V}{d\phi}$. An overdot denotes differentiation with
respect to the comoving time, $t$.

\section{Induced Gravity with Non-Minimally Coupled Brane-Scalar
Field}

The action of the DGP scenario in the presence of a non-minimally
coupled scalar field on the brane can be written as follows
\begin{equation}
S=\int d^{5}x\frac{m^{3}_{4}}{2}\sqrt{-g}{\cal R}+\Bigg[\int
d^{4}x\sqrt{-q}\bigg(\frac{m_{3}^{2}}{2}\alpha(\phi)
R[q]-\frac{1}{2} q^{\mu\nu} \nabla_{\mu}\phi\nabla_{\nu}\phi
-V(\phi) + m^{3}_{4}\overline{K}+ {\cal{L}}_{m}\bigg)\Bigg]_{y=0},
\end{equation}
where we have included a general non-minimal coupling $\alpha(\phi)$
\, in the brane part of the action\footnote{ For an interesting
discussion on the importance of non-minimal coupling and possible
schemes to incorporate it in the formulation of scalar-tensor
gravity see [31].}.\, $y$ is coordinate of fifth dimension and we
assume brane is located at $y=0$.\, $g_{AB}$ is five dimensional
bulk metric with Ricci scalar ${\cal{R}}$, while $q_{\mu\nu}$ is
induced metric on the brane with induced Ricci scalar $R$.\,
$g_{AB}$ and $q_{\mu\nu}$ are related via
$q_{\mu\nu}={\delta_{\mu}}^{A}{\delta_{\nu}}^{B}g_{AB}$.\,
$\overline{K}$ is trace of the mean extrinsic curvature of the brane
defined as
\begin{equation}
\overline{K}_{\mu\nu}=\frac{1}{2}\,\,\lim_{\epsilon\rightarrow
0}\bigg(\Big[K_{\mu\nu}\Big]_{y=-\epsilon}+
\Big[K_{\mu\nu}\Big]_{y=+\epsilon}\bigg),
\end{equation}
and corresponding term in the action is York-Gibbons-Hawking
term[32] (see also [20]). The ordinary matter part of the action is
shown by Lagrangian ${\cal{L}}_{m}\equiv
{\cal{L}}_{m}(q_{\mu\nu},\psi)$ where $\psi$ is matter field and
corresponding energy-momentum tensor is
\begin{equation}
T_{\mu\nu}=-2\frac{\delta{\cal{L}}_{m}}{\delta
q^{\mu\nu}}+q_{\mu\nu}{\cal{L}}_{m}.
\end{equation}
The pure scalar field Lagrangian,\, ${\cal{L}}_{\phi}=-\frac{1}{2}
q^{\mu\nu} \nabla_{\mu}\phi\nabla_{\nu}\phi -V(\phi)$,\,\,  yields
the following energy-momentum tensor
\begin{equation}
 \tau_{\mu\nu}=\nabla_\mu\phi\nabla_\nu\phi-\frac{1}{2}q_{\mu\nu}(\nabla\phi)^2
-q_{\mu\nu}V(\phi)
\end{equation}
The Bulk-brane Einstein's equations calculated from action (1) are
given by
$$m^{3}_{4}\left({\cal R}_{AB}-\frac{1}{2}g_{AB}{\cal
R}\right)+$$
\begin{equation}
m^{2}_{3}{\delta_{A}}^{\mu}{\delta_{B}}^{\nu}\bigg[\alpha(\phi)\left(R_{\mu\nu}-
\frac{1}{2}q_{\mu\nu}R\right)-\nabla_{\mu}\nabla_{\nu}\alpha(\phi)+q_{\mu\nu}\Box^{(4)}\alpha(\phi)\bigg]\delta(y)
={\delta_{A}}^{\mu}{\delta_{B}}^{\nu}\Upsilon_{\mu\nu}\delta(y),
\end{equation}
where $\Box^{(4)}$ is 4-dimensional(brane) d'Alembertian and
$\Upsilon_{\mu\nu}=T_{\mu\nu}+\tau_{\mu\nu}$\,. This relation can be
written as follows
\begin{equation}
m^{3}_{4}\left({\cal R}_{AB}-\frac{1}{2}g_{AB}{\cal R}\right)+
m^{2}_{3}\alpha(\phi){\delta_{A}}^{\mu}{\delta_{B}}^{\nu}\left(R_{\mu\nu}-
\frac{1}{2}q_{\mu\nu}R\right)\delta(y)=
{\delta_{A}}^{\mu}{\delta_{B}}^{\nu}{\cal{T}}_{\mu\nu}\delta(y)
\end{equation}
where ${\cal{T}}_{\mu\nu}$ is total energy-momentum on the brane
defined as follows
\begin{equation}
{\cal{T}}_{\mu\nu}=m^{2}_{3}\nabla_{\mu}\nabla_{\nu}\alpha(\phi)-m^{2}_{3}
q_{\mu\nu}\Box^{(4)}\alpha(\phi)+\Upsilon_{\mu\nu},
\end{equation}
From (6) we find
\begin{equation}
G_{AB}={\cal R}_{AB}-\frac{1}{2}g_{AB}{\cal R}=0
\end{equation}
and
\begin{equation}
G_{\mu\nu}=\left(R_{\mu\nu}-
\frac{1}{2}q_{\mu\nu}R\right)=\frac{{\cal
T}_{\mu\nu}}{m^{2}_{3}\alpha(\phi)}
\end{equation}
for bulk and brane respectively. The corresponding junction
conditions relating the extrinsic curvature to the energy-momentum
tensor of the brane, have the following form
\begin{equation}
\lim_{\epsilon\rightarrow+0}\Big[K_{\mu\nu}\Big]^{y=+\epsilon}_{y=-\epsilon}
=\frac{1}{m_{4}^{3}}\bigg[{\cal{T}}_{\mu\nu}-\frac{1}{3}q_{\mu\nu}q^{\alpha\beta}
{\cal {T}}_{\alpha\beta}\bigg]_{y=0}
-\frac{m^{2}_{3}\alpha(\phi)}{m^{3}_{4}}\bigg[R_{\mu\nu}-
\frac{1}{6}q_{\mu\nu}q^{\alpha\beta}R_{\alpha\beta}\bigg]_{y=0}.
\end{equation}

\section{Embedding of FRW Cosmology}
We start with the following line element to derive cosmological
dynamics of our model,
\begin{equation}
ds^{2}=q_{\mu\nu}dx^{\mu}dx^{\nu}+b^{2}(y,t)dy^{2}=-n^{2}(y,t)dt^{2}+
a^{2}(y,t)\gamma_{ij}dx^{i}dx^{j}+b^{2}(y,t)dy^{2}.
\end{equation}
In this relation $\gamma_{ij}$ is a maximally symmetric
3-dimensional metric defined as
\begin{equation}
\gamma_{ij}=\delta_{ij}+k\frac{x_{i}x_{j}}{1-kr^{2}}
\end{equation}
where $k=-1,0,1$ parameterizes the spatial curvature and
$r^2=x_{i}x^{i}$. We assume that scalar field $\phi$ depends only on
the proper cosmic time of the brane. Choosing gauge $b^{2}(y,t)=1$
in Gaussian normal coordinates, the field equations in the bulk are
given by (8) with the following Einstein's tensor components
\begin{equation}
G_{00}=3n^{2} \Big(\frac{\dot{a}^2}{n^2a^2}-\frac{{a'}^2}{a^2}
-\frac{a''}{a}+\frac{k}{a^2}\Big),
\end{equation}
\begin{equation}
G_{ij}=\gamma_{ij}a^{2}\bigg[\Big(\frac{{a'}^2}{a^2}
-\frac{\dot{a}^2}{n^2a^2}-\frac{k}{a^2}\Big)
+2\Big(\frac{a''}{a}+\frac{n'a'}{na}
-\frac{\ddot{a}}{n^2a}+\frac{\dot{n}\dot{a}}{n^3a}+\frac{n''}{2n}\Big)\Bigg]
\end{equation}
\begin{equation}
G_{0y}=3\Big( \frac{n'}{n}\frac{\dot{a}}{a}-\frac{\dot{a}'}{a}
\Big),
\end{equation}
\begin{equation}
G_{yy}=3\Big(\frac{{a'}^2}{a^2}-\frac{\dot{a}^2}{n^2a^2}
-\frac{k}{a^2} +\frac{n'a'}{na}+\frac{\dot{n}\dot{a}}{n^3a}
-\frac{\ddot{a}}{n^2a}\Big).
\end{equation}
The field equations on the brane are given by the following
equations
\begin{equation}
G^{(3)}_{00}=3n^{2}\left(\frac{\dot{a}^{2}}{n^{2}a^{2}}+
\frac{k}{a^{2}}\right)=\frac{2}{m^{2}_{3}\alpha(\phi)}{\cal
{T}}_{00},
\end{equation}
\begin{equation}
G^{(3)}_{ij}=\gamma_{ij}\bigg[2\bigg(\frac{\dot{n}\dot{a}}{n^{3}a}-
\frac{\ddot{a}}{n^{2}a}\bigg)-\bigg(\frac{\dot{a}^{2}}{n^{2}a^{2}}+
\frac{k}{a^{2}}\bigg)\bigg]=\frac{2}{m^{2}_{3}\alpha(\phi)}{\cal
T}_{ij},
\end{equation}
and scalar field evolution equation
\begin{equation}
\ddot{\phi}+\bigg(3\frac{\dot a}{a}-\frac{\dot
n}{n}\bigg)\dot{\phi}+n^{2}\frac{dV}{d\phi}-\frac{m_{3}^2}{2}n^{2}\alpha'R[q]=0,
\end{equation}
where Ricci scalar on the brane is given by
\begin{equation}
R=3\frac{k}{a^2}+\frac{1}{n^{2}}\bigg[6\frac{\ddot{a}}{a}+6\Big(\frac{\dot{a}}{a}\Big)^{2}-6\frac{\dot{a}}{a}\frac{\dot{n}}{n}\bigg].
\end{equation}
The other important equation is the continuity equation on the
brane. Suppose that ordinary matter on the brane has an ideal fluid
form, $T_{\mu\nu}=(\rho+p)u_{\mu}u_{\nu}+pq_{\mu\nu}$. Since
$K_{tt}=nn'$ and $K_{rr}=-aa'$, equation (10) gives the following
matching conditions
\begin{equation}
\lim_{\epsilon\rightarrow+0}\left[\partial_{y}a\right]^{y=
+\epsilon}_{y=-\epsilon}(t)=\frac{m^{2}_{3}}{m^{3}_{4}}\Bigg[\alpha(\phi)
\bigg(\frac{\dot{a}^{2}}{n^{2}a}+ \frac{k}{a}\bigg)\Bigg]_{y=0}
-\Bigg[\frac{(\rho+\rho_{\phi})a}{3 m^{3}_{4}}\Bigg]_{y=0}.
\end{equation}
$$\lim_{\epsilon\rightarrow+0}\left[\partial_{y}n\right]^{y=
+\epsilon}_{y=-\epsilon}(t)=\frac{m^{2}_{3}}{m^{3}_{4}}(2n)\Bigg[\alpha(\phi)\Big(\frac{\ddot{a}}{n^{2}a}-\frac{\dot{a}^{2}}{2n^{2}a^2}-
\frac{\dot{n}\dot{a}}{n^{3}a}-\frac{k}{2a^{2}}\Big) \Bigg]_{y=0}+$$
\begin{equation}
\frac{n}{3m_{4}^{3}}\Bigg[2(\rho+\rho_{\phi})+3(p+p_{\phi})\Bigg]_{y=0}
\end{equation}
where energy density and pressure of non-minimally coupled scalar
field are given as follows
\begin{equation}
\rho_{\phi}=\left[\frac{1}{2}\dot{\phi}^{2}+n^{2}V(\phi)-6\alpha'H\dot{\phi}\right]_{y=0},
\end{equation}
\begin{equation}
p_{\phi}=\left[\frac{1}{2n^{2}}\dot{\phi}^{2}-V(\phi)+
\frac{2\alpha'}{n^2}\Big(\ddot{\phi}-\frac{\dot{n}}{n}\dot{\phi}\Big)+
4\alpha'\frac{H}{n^{2}}\dot{\phi}+\frac{2\alpha''}{n^2}\dot{\phi}^2
\right]_{y=0},
\end{equation}
and $H=\frac{\dot{a}}{a}$ is Hubble parameter. Note that part of the
effect of non-minimal coupling of the field $\phi$ is hidden in the
definition of the effective energy density and pressure which both
include non-minimal terms. Now using (15) since in the bulk
$G_{00}=0$, we find
\begin{equation}
\lim_{\epsilon\to+0}\bigg[\frac{n'}{n}\bigg]_{y=-\epsilon}^{y=+\epsilon}=
\bigg[\frac{\dot{a}'}{\dot{a}}\bigg]_{y=-\epsilon}^{y=+\epsilon}
\end{equation}
using relations (21) and (22) we find the following relation for
conservation of energy on the brane
\begin{equation}
\dot{\rho}+\dot{\rho}_{\phi}+3H\Big(\rho+\rho_{\phi}+p+p_{\phi}\Big)=6\alpha'\dot\phi
\Big(H^2+\frac{k}{a^2}\Big).
\end{equation}
Thus the non-minimal coupling of the scalar field to the Ricci
curvature on the brane through $\alpha(\phi)$ leads to the
non-conservation of the effective energy density.

To obtain the cosmological dynamics, we set $n(0,t)=1$. With this
gauge condition we recover usual time on the brane via
transformation $t=\int^{t}n(0,\eta)d\eta$. In this situation, our
basic dynamical variable is only $a(y,t)$ since $n(y,t)$ now is
given by
\begin{equation}
n(y,t)=\frac{\dot{a}(y,t)}{\dot{a}(0,t)}.
\end{equation}
where $H=\frac{\dot{a}(0,t)}{a(0,t)}$ is Hubble parameter on the
brane. Now we can write the basic set of cosmological equations for
a FRW brane in the presence of a non-minimally coupled scalar field.
The first of these equations is given by matching condition
\begin{equation}
\lim_{\epsilon\rightarrow+0}\left[\partial_{y}a\right]^{y=
+\epsilon}_{y=-\epsilon}(t)=\frac{m^{2}_{3}}{m^{3}_{4}}\Bigg[\alpha(\phi)
\bigg(\frac{\dot{a}^{2}}{n^{2}a}+ \frac{k}{a}\bigg)\Bigg]_{y=0}
-\Bigg[\frac{(\rho+\rho_{\phi})a}{3 m^{3}_{4}}\Bigg]_{y=0}.
\end{equation}
Insertion of\,\,$\frac{n'}{n}=\frac{\dot{a}'}{\dot{a}}$\,\, into
equations (13) and (16) yields the Bin\'{e}truy {\it et al}\,\,[14]
integral
\begin{equation}
{\cal
I}^{+}=\bigg[\Big(\frac{\dot{a}^{2}}{n^2}-a'^{2}+k\Big)a^{2}\bigg]_{y>0},
\end{equation}
and
\begin{equation}
{\cal
I}^{-}=\bigg[\Big(\frac{\dot{a}^{2}}{n^2}-a'^{2}+k\Big)a^{2}\bigg]_{y<0},
\end{equation}
which are constant and if $a'$ is continuous on the brane then
${\cal I}^{+}={\cal I}^{-}$. These equations along with scalar field
equation
\begin{equation}
\ddot{\phi}+\bigg(3\frac{\dot a}{a}-\frac{\dot
n}{n}\bigg)\dot{\phi}+n^{2}\frac{dV}{d\phi}-n^{2}\frac{d\alpha}{d\phi}R[q]=0,
\end{equation}
and
\begin{equation}
n(y,t)=\frac{\dot{a}(y,t)}{\dot{a}(0,t)}.
\end{equation}
constitute the basic dynamical equations of our model. In the
absence of transverse momentum, $\Upsilon_{0y}=0$, one has ${\cal
I}^{+}={\cal I}^{-}$. In fact ${\cal I}^{\pm}$ can be considered as
initial conditions and these quantities reflect the symmetry across
the brane.  We first consider the case ${\cal I}^{+}={\cal I}^{-}$
in which follows. Our cosmological equations on the brane now take
the following forms(note that $n(0,t)=1$)
\begin{equation}
\frac{\dot{a}^{2}(0,t)+k}{a^{2}(0,t)}=
\frac{(\rho+\rho_{\phi})}{3m^{2}_{3}\alpha(\phi)},
\end{equation}

\begin{equation}
\ddot{\phi}+3\frac{\dot{a}(0,t)}{a(0,t)}\dot{\phi}+\frac{dV(\phi)}{d\phi}
=\frac{d\alpha}{d\phi}R[q],
\end{equation}
\begin{equation}
{\cal I}=\Big[\dot{a}^{2}(0,t)-a'^{2}(y,t)+k\Big]a^{2}(y,t)
\end{equation}
\begin{equation}
n(y,t)=\frac{\dot{a}(y,t)}{\dot{a}(0,t)}.
\end{equation}
Using equation (35), the scale factor is calculated as follows
\begin{equation}
a^{2}(y,t)=a^{2}(0,t)+ \Big[\dot{a}^{2}(0,t)+k\Big]y^{2}
+2\bigg[\Big(\dot{a}^{2}(0,t)+k\Big) a^{2}(0,t)-{\cal
I}\bigg]^{\frac{1}{2}}y
\end{equation}
and therefore $n(y,t)$ is given by equation (36);
$$n(y,t)=\Bigg(a(0,t)+\ddot{a}(0,t)y^{2}+a(0,t)\frac{a(0,t)\ddot{a}(0,t)+
\dot{a}^{2}(0,t)+k}{\sqrt{\Big(\dot{a}^{2}(0,t)+k\Big)a^{2}(0,t)-{\cal
{I}}}}y\Bigg)$$
\begin{equation}
\times \Bigg[a^{2}(0,t)+ \Big[\dot{a}^{2}(0,t)+k\Big]y^{2}
+2\Big[\Big(\dot{a}^{2}(0,t)+k\Big) a^{2}(0,t)-{\cal
I}\Big]^{\frac{1}{2}}y\Bigg]^{\frac{-1}{2}}
\end{equation}
So, the component of 5-dimensional metric (11) are determined. If we
set initial conditions in such a way that ${\cal I}= 0$, we find the
following simple equations for cosmological dynamics
\begin{equation}
 a(y,t)=a(0,t)+\Big[\dot{a}^{2}(0,t)+k\Big]^{\frac{1}{2}}y,
\end{equation}
\begin{equation}
n(y,t)=1+\frac{\ddot{a}(0,t)}{\sqrt{\dot{a}^{2}(0,t)+k}}y.
\end{equation}
Therefore, our model allows for an embedding of the standard
Friedmann cosmology in the sense that the cosmological evolution of
the background metric on the brane can be described by the standard
Friedmann equation.

So far we have discussed the case ${\cal I}^{+}={\cal I}^{-}$ with a
continuous warp factor across the brane. In the case of ${\cal
I}^{+}\neq{\cal I}^{-}$, there cannot be any symmetry across the
brane. In this case the basic set of dynamical equations is provided
by equations (28), (29), (30) plus the non-conservation of the
effective energy density given by (26). In this case, evolution of
the scale factor on the brane is given by elimination of
$a'(y\longrightarrow\pm0,t)$ from the following generalized
Friedmann equation

$$\pm\bigg[\dot{a}^{2}(0,t)+k-a^{-2}(0,t){\cal
I}^{+}\bigg]^{\frac{1}{2}}\mp\bigg[\dot{a}^{2}(0,t)+k-a^{-2}(0,t){\cal
I}^{-}\bigg]^{\frac{1}{2}}$$
\begin{equation}
\quad\quad\quad\quad\quad\quad=\alpha(\phi)\frac{m_{3}^{2}}{m_{4}^{3}}
\bigg(\frac{\dot{a}^{2}(0,t)+k}{a(0,t)}\bigg)-
\frac{(\rho+\rho_{\phi})a(0,t)}{3m^{3}_{4}}.
\end{equation}
This is the most general form of modified Friedmann equation for our
non-minimal framework. After determination of $a(0,t)$, since ${\cal
I}^{\pm}$ are constants, \, $a(y,t)$ can be calculated from (35).
This is the full dynamics of the system. Note that in the case where
the right hand side of equation (41) is negative, at least one sign
in left hand side should be negative depending on initial
conditions. However, the dynamics of the problem does not require
symmetry across the brane. Therefore, we have shown the possibility
of embedding of FRW cosmology in DGP scenario with a 4D
non-minimally coupled scalar field on the brane and equation (41) is
the most general form of FRW equation in this embedding. This
relation for the case with ${\cal I}^{+}={\cal I}^{-}\equiv{\cal I}$
and a discontinuous warp factor across the $Z_{2}$ symmetric brane
leads to the well-known generalization of Friedmann equation in dark
energy formalism. To show this feature, we define for simplicity,
$$ x\equiv H^{2}+\frac{k}{a^{2}},$$
$$b\equiv\rho+\rho_{\phi},$$
$$y\equiv\alpha(\phi)m_{3}^{2},$$
and
$$z\equiv m_{4}^{3}.$$
With these definitions, equation (41) (with upper sign for
instance), transforms to the following form
\begin{equation}
\bigg(x-\frac{{\cal
I}^{+}}{a^4}\bigg)^{\frac{1}{2}}+\bigg(x-\frac{{\cal
I}^{-}}{a^4}\bigg)^{\frac{1}{2}}=\frac{y}{z}x-\frac{b}{3z}.
\end{equation}
Solving this equation for $x$ (with ${\cal I}^{+}={\cal
I}^{-}\equiv{\cal I}$) gives the following result
\begin{equation}
x=\frac{\frac{by}{3z^2}+2\pm\sqrt{\Big(\frac{by}{3z^2}+2\Big)^{2}
-\frac{y^2}{z^{2}}\Big(\frac{b^2}{9z^2}+\frac{4{\cal
I}}{a^4}\Big)}}{\frac{y^2}{z^2}}.
\end{equation}
A little algebraic manipulation gives
\begin{equation}
x=\frac{1}{3y}\Bigg[b+\frac{6z^2}{y}\pm\frac{6z^2}{y}
\sqrt{1+\frac{by}{3z^2}-\frac{{\cal I}y^2}{a^{4}z^{2}}}\Bigg].
\end{equation}
Considering both plus and minus signs in equation (41) and using
original quantities we obtain
\begin{equation}
H^{2}+\frac{k}{a^2}=\frac{1}{3m_{3}^{2}\alpha(\phi)}\bigg(\rho+\rho_{\phi}+\rho_{0}\Big[1+\varepsilon
\sqrt{1+\frac{2}{\rho_{0}}\Big[\rho+\rho_{\phi}-m_{3}^{2}\alpha(\phi)\frac{{{\cal{E}}_{0}}}{a^{4}}\Big]}\,\,\bigg).
\end{equation}
where
$\rho_{0}\equiv\frac{6z^2}{y}=\frac{6m_{4}^{6}}{m_{3}^{2}\alpha(\phi)}$,
\, $\varepsilon=\pm1$ \, shows the possibility of existence of two
different branches of FRW equation and \, ${\cal{E}}_{0}=3{\cal
I}$\,\, is a constant. This analysis shows the consistency of our
formalism with dark-radiation formalism presented in [23,30]. In the
high energy regime where $\frac{\rho+\rho_{\phi}}{\rho_{0}}\gg 1$,
we find
\begin{equation}
H^{2}+\frac{k}{a^2}\approx\frac{1}{3m_{3}^{2}\alpha(\phi)}\Big(\rho+\rho_{\phi}+\varepsilon\sqrt{2(\rho+\rho_{\phi})\rho_{0}}\,\,\Big)
\end{equation}
which describes a four dimensional gravity with a small correction.
Neglecting this small correction, the resulting equation is exactly
the same as equation (33). In the low energy regime where
$\frac{\rho+\rho_{\phi}}{\rho_{0}}\ll 1$, we find
\begin{equation}
H^{2}+\frac{k}{a^2}\approx\frac{1}{3m_{3}^{2}\alpha(\phi)}
\bigg[(1+\varepsilon)(\rho+\rho_{\phi})+(1+\varepsilon)\rho_{0}
-\frac{\varepsilon}{4}\frac{(\rho+\rho_{\phi})^{2}}{\rho_{0}}\bigg].
\end{equation}
For $\varepsilon=+1$ this equation describes a four-dimensional
gravity, while for $\varepsilon=-1$ we have a five-dimensional
gravity.

\section{Summary and Conclusions}
In this paper we have considered the DGP model with a non-minimally
coupled scalar field on the brane. The introduction of non-minimal
coupling is not just a matter of taste; it is forced upon us in many
situations of physical and cosmological interests such as quantum
corrections to the scalar field theory and its renormalizability in
curved spacetime. In the spirit of DGP inspired gravity, we have
studied the effect of an induced gravity term which is an arbitrary
function of a scalar field on the brane. We have presented
four-dimensional equations on a DGP brane with a scalar field
non-minimally coupled to the induced Ricci curvature, embedded in a
five-dimensional Minkowski bulk. This is an extension to a
braneworld context of scalar-tensor (Brans-Dicke) gravity. Our model
does not describe a special dynamics for late time acceleration
beyond the standard DGP scenario. In other words, as original DGP
scenario and its minimal extension which support late-time
acceleration of the universe[20,29], it is easy to show that our
model contains this late time acceleration for a suitable range of
non-minimal coupling. However, with non-minimally coupled scalar
field on the brane, generally it is harder to achieve accelerated
expansion[31,33]. Only with suitable choice of non-minimal coupling
and scalar field potential one can achieve accelerated expansion in
this non-minimal model. The main ingredient of our analysis lies in
the fact that DGP model allows for an embedding of the standard
Friedmann cosmology in the sense that the cosmological evolution of
the background metric on the brane can entirely be described by the
standard Friedmann equation plus {\it total} energy conservation on
the brane and the dynamics of the problem does not require symmetry
across the brane. Our general framework applied to a
$Z_{2}$-symmetric brane gives the
well-known result of dark energy formulation.\\

{\bf Acknowledgement}\\
It is a pleasure to appreciate members of the Centre for Particle
Theory at Durham University, specially Professor Ruth Gregory for
hospitality. I would like also to appreciate referee for his/her
important contributions in this work.

\end{document}